\documentclass[a4paper,11pt]{article}
\usepackage{pos}

\newcommand{\Ddag}{\ensuremath{D^{\dag}} }
\newcommand{\cD}{\ensuremath{\mathcal D} }
\newcommand{\cDbar}{\ensuremath{\overline{\mathcal D}} }

\newcommand{\Ibb}{\ensuremath{\mathbb I} }
\newcommand{\cN}{\ensuremath{\mathcal N} }
\newcommand{\cO}{\ensuremath{\mathcal O} }
\newcommand{\cP}{\ensuremath{\mathcal P} }
\newcommand{\cQ}{\ensuremath{\mathcal Q} }
\newcommand{\cU}{\ensuremath{\mathcal U} }
\newcommand{\cUbar}{\ensuremath{\overline{\mathcal U}} }
\newcommand{\ga}{\ensuremath{\gamma} }
\newcommand{\gaEff}{\ensuremath{\ga_{\text{eff}}} }
\newcommand{\de}{\ensuremath{\delta} }
\newcommand{\vareps}{\ensuremath{\varepsilon} }
\newcommand{\De}{\ensuremath{\Delta} }
\newcommand{\la}{\ensuremath{\lambda} }
\newcommand{\lalat}{\ensuremath{\la_{\text{lat}}} }
\newcommand{\muhat}{\ensuremath{\widehat\mu} }
\newcommand{\om}{\ensuremath{\omega} }
\newcommand{\Om}{\ensuremath{\Omega} }

\newcommand{\spacer}{\textcolor{white}{(}}
\newcommand{\SO}[1]{\ensuremath{\text{SO(}#1\text{)}} }
\newcommand{\SU}[1]{\ensuremath{\text{SU(}#1\text{)}} }
\newcommand{\U}[1]{\ensuremath{\text{U(}#1\text{)}} }
\newcommand{\Uone}{\ensuremath{\text{U(1)}} }
\DeclareMathOperator{\Tr}{Tr}
\newcommand{\lra}{\ensuremath{\longrightarrow} }
\newcommand{\llra}{\ensuremath{\longleftrightarrow} }
\newcommand{\vev}[1]{\ensuremath{\left\langle #1 \right\rangle} }
\newcommand{\pderiv}[2]{\ensuremath{\frac{\partial #1}{\partial #2}} }
\newcommand{\eq}[1]{Eq.~\ref{#1}}
\newcommand{\fig}[1]{Fig.~\ref{#1}}
\newcommand{\secref}[1]{Section~\ref{#1}}
\newcommand{\refcite}[1]{Ref.~\cite{#1}}

\title{Exploring conformality \\ \hfill in lattice $\cN = 4$ supersymmetric Yang--Mills}
\ShortTitle{Exploring conformality in lattice $\cN = 4$ supersymmetric Yang--Mills}

\author*{David Schaich}
\affiliation{Department of Mathematical Sciences, University of Liverpool, Liverpool L69 7ZL, United Kingdom}
\emailAdd{david.schaich@liverpool.ac.uk}

\FullConference{39th International Symposium on Lattice Field Theory \\ 8--13 August 2022 \\ Rheinische Friedrich-Wilhelms-Universit{\"a}t Bonn, Germany}

\abstract{ 
  Maximally supersymmetric Yang--Mills theory ($\cN = 4$ SYM) is conformal for any value of the coupling.
  Lattice regularization breaks conformality through the introduction of a non-zero lattice spacing and a finite lattice volume.
  This proceedings presents ongoing numerical computations of conformal scaling dimensions in lattice $\cN = 4$ SYM, based on a lattice formulation that exactly preserves a supersymmetry sub-algebra at non-zero lattice spacing.
  The main targets are the non-trivial anomalous dimension of the Konishi operator, as well as a mass anomalous dimension extracted from the eigenvalue mode number of the fermion operator.
  The latter is expected to vanish in the conformal continuum theory, providing insight into the interplay of lattice discretization and conformality.
}

\begin{document}
\maketitle

\section{Introduction} 
Maximally supersymmetric Yang--Mills theory ($\cN = 4$ SYM) is widely studied in theoretical physics.
It is arguably the simplest non-trivial quantum field theory in four dimensions, especially in the large-$N$ planar limit of its SU($N$) gauge group, thanks to its many symmetries.
It plays a key role in holographic duality, provided early insight into S-duality, and continues to inform modern analyses of scattering amplitudes.

Lattice regularization of $\cN = 4$ SYM is an active area of research that provides both a non-perturbative definition of the theory as well as a way to numerically predict its behavior from first principles, even at strong coupling and away from the planar limit.
See \refcite{Schaich:2022xgy} for a recent review.
Lattice formulations of $\cN = 4$ SYM necessarily break symmetries of the continuum theory, and most of the recent activity has relied on an approach that preserves a closed supersymmetry sub-algebra at non-zero lattice spacing $a > 0$~\cite{Schaich:2022xgy, Catterall:2009it}.
Despite preserving only a single one of the 16 supersymmetries, this single exact supersymmetry significantly simplifies the lattice theory, making it possible to ensure the recovery of the other 15 in the $a \to 0$ continuum limit.

This proceedings focuses on another challenge in lattice studies of $\cN = 4$ SYM.
Although the continuum theory is conformal for every value of the 't~Hooft coupling $\la = N g^2$, numerical lattice field theory calculations require both a non-zero lattice spacing corresponding to a UV cutoff $\sim 1 / a$, as well as a finite lattice volume $(L\!\cdot\!a)^4$ introducing an IR cutoff.
Both of these explicitly break conformal scale invariance, complicating lattice analyses of the spectrum of $\la$-dependent conformal scaling dimensions, which is key information about the theory.

After a brief review of lattice $\cN = 4$ SYM in the next section, I will summarize recent and ongoing numerical investigations of two scaling dimensions of particular interest.
First, \secref{sec:mode} presents results from \refcite{Bergner:2021ffz} for a mass anomalous dimension obtained by analyzing the eigenmode number of the fermion operator.
This anomalous dimension is expected to vanish, $\ga_* = 0$, for all values of $\la$, which allows numerical results to reveal the effects of breaking conformality (and supersymmetry).
Second, in \secref{sec:Konishi} I show preliminary results from ongoing lattice studies of the non-trivial Konishi scaling dimension $\De_K(\la)$.
Section~\ref{sec:conc} concludes with a brief discussion of next steps for this line of research.

\section{Lattice $\cN = 4$ SYM in a nutshell} 
The lattice formulation used in this work is based on so-called topological twisting, which organizes the 16 supercharges of the theory into integer-spin representations of the twisted rotation group $\SO{4}_{\text{tw}} \equiv \mbox{diag}\left[\SO{4}_{\textrm{euc}} \otimes \SO{4}_R\right]$, where $\SO{4}_{\textrm{euc}}$ is the Lorentz group Wick-rotated to euclidean space-time and $\SO{4}_R$ is a subgroup of the SO(6) R-symmetry.
The twisted-scalar supersymmetry \cQ is nilpotent, preserving the sub-algebra $\{\cQ, \cQ\} = 0$ even at non-zero lattice spacing where the other 15 supersymmetries are broken by the lattice discretization of space-time.

The lattice theory is formulated on the $A_4^*$ lattice, which complicates the relation between the lattice 't~Hooft coupling \lalat the continuum $\la$~\cite{Catterall:2014vka}.
The lattice action features the same $\cQ$-exact and $\cQ$-closed terms as the twisted continuum theory:
\begin{equation}
  \label{eq:action}
  \begin{split}
    S & = \frac{N}{4\lalat} \sum_n \Tr\left[\cQ \left(\chi_{ab}(n)\cD_a^{(+)}\cU_b(n) + \eta(n) \cDbar_a^{(-)}\cU_a(n) - \frac{1}{2}\eta(n) d(n) \right)\right] \\
      & \qquad -\frac{N}{16\lalat} \sum_n \Tr\left[\vareps_{abcde}\ \chi_{de}(n + \muhat_a + \muhat_b + \muhat_c) \cDbar_c^{(-)} \chi_{ab}(n)\right].
  \end{split}
\end{equation}
Here $\eta$, $\psi_a$ and $\chi_{ab} = -\chi_{ba}$ are $1$-, $5$- and $10$-component fermions respectively associated with the sites, links and oriented plaquettes of the $A_4^*$ lattice.
The five-component complexified gauge links $\cU_a$ and $\cUbar_a$ contain both the gauge and scalar fields, and also appear in the finite-difference operators $\cD_a^{(+)}$ and $\cDbar_a^{(-)}$.

The complexified gauge links lead to $\U{N} = \SU{N}\times \Uone$ gauge invariance.
In order to carry out numerical calculations, flat directions in both the SU($N$) and U(1) sectors need to be regulated.
We achieve this by adding two deformations to \eq{eq:action}, following \refcite{Catterall:2015ira}.\footnote{There is ongoing exploration of alternative lattice actions, including \refcite{Catterall:2020lsi}.}
First, the double-trace scalar potential
\begin{equation}
  \label{eq:pot}
  S_{\text{scalar}} = \frac{N}{4\lalat} \mu^2 \sum_n \sum_a \left(\frac{1}{N} \Tr\left[\cU_a(n) \cUbar_a(n)\right] - 1\right)^2
\end{equation}
regulates the SU($N$) flat directions while softly breaking the \cQ supersymmetry.
The second deformation is $\cQ$-exact, and replaces the term
\begin{equation}
  \label{eq:det}
  \cQ \left(\eta(n) \cDbar_a^{(-)}\cU_a(n)\right) \lra \cQ \left(\eta(n) \left[\cDbar_a^{(-)}\cU_a(n) + G\sum_{a \neq b} \left(\det\cP_{ab}(n) - 1\right) \Ibb_{N}\right]\right)
\end{equation}
in \eq{eq:action}.
This picks out the U(1) sector through the determinant of the plaquette oriented in the $a$--$b$ plane, $\cP_{ab}(n)$, which is an $N\times N$ matrix at each lattice site $n$.
The tunable parameter $\mu$ needs to be sent to zero in the continuum limit in order to restore supersymmetry, while the U(1) deformation involving $G$ can be expected to decouple.

Using parallel software that we present in \refcite{Schaich:2014pda} and make publicly available through {\tt\href{https://github.com/daschaich/susy}{github.com/daschaich/susy}}, we have used the rational hybrid Monte Carlo (RHMC) algorithm to generate many ensembles of field configurations, a subset of which we have analyzed for \refcite{Bergner:2021ffz} and this proceedings.
In the time since \refcite{Schaich:2014pda} appeared, we have extended the software to implement the improved action discussed above, as well as the stochastic estimation of the eigenmode number discussed in the next section, which rescales some of the fermion field components in order to put the fermion operator into its most symmetric form.
An updated publication will soon present these improvements.
In the meantime \refcite{Catterall:2015ira} provides more information about the improved action while \refcite{Bergner:2021ffz} discusses the fermion rescaling and stochastic eigenmode calculation in more detail.

\section{\label{sec:mode}Mass anomalous dimension from fermion eigenmode number} 
Building on the early work of \refcite{Weir:2013zua}, the recent study \refcite{Bergner:2021ffz} uses efficient modern techniques to evaluate the eigenmode number of the fermion operator, which is related to a mass anomalous dimension $\ga_*$~\cite{Patella:2012da, Cheng:2013eu, Cheng:2013bca, Fodor:2014zca}.
We start with the spectral density of $\Ddag D$, where $D$ is the massless lattice fermion operator:
\begin{equation}
  \rho(\om^2) = \frac{1}{V} \sum_k \vev{\de(\om^2 - \la_k^2)}.
\end{equation}
Here the eigenvalues $\la_k$ of $D$ should not be confused with the 't~Hooft coupling $\la$.
The eigenmode number $\nu(\Om^2)$ is the number of eigenvalues $\la_k^2$ of the non-negative operator $\Ddag D$ that are smaller than $\Om^2$:
\begin{equation}
  \label{eq:mode}
  \nu(\Om^2) = \int_0^{\Om^2} \rho(\om^2) \mathop{d\om^2} \propto \left(\Om^2\right)^{2 / (1 + \ga_*)}.
\end{equation}
Note that the lattice $\cN = 4$ SYM fermion operator $D$ is antisymmetric,
\begin{equation*}
  \Psi^T D \Psi = \chi_{ab} \cD_{[a}^{(+)} \psi_{b]}^{\spacer} + \eta \cDbar_a^{(-)} \psi_a + \frac{1}{2}\vareps_{abcde} \chi_{ab} \cDbar_c^{(-)} \chi_{de},
\end{equation*}
which ensures that the eigenvalues of $D$ come in $\pm \la_k$ pairs.

In order to efficiently evaluate the spectral density and eigenmode number over the full spectral range of each lattice ensemble, we adopt the method~\cite{Fodor:2016hke, Bergner:2016hip} of stochastically estimating the Chebyshev expansion
\begin{equation}
  \rho_r(x) \approx \sum_{n = 0}^P \frac{2 - \de_{n0}}{\pi\sqrt{1 - x^2}} c_n T_n(x).
\end{equation}
For lattice ensembles with gauge groups U(2), U(3), U(4) and lattice volumes up to $16^4$, we retain $5000 \leq P \leq 10000$ terms in the Chebyshev expansion.
In \refcite{Bergner:2021ffz} we checked selected results against both direct iterative evaluation of the low-lying eigenvalues as well as stochastic projection computations.
Both of those methods are much more computationally expensive than this stochastic Chebyshev estimation.

\begin{figure}[tbp]
  \centering
  \includegraphics[width=0.9\linewidth]{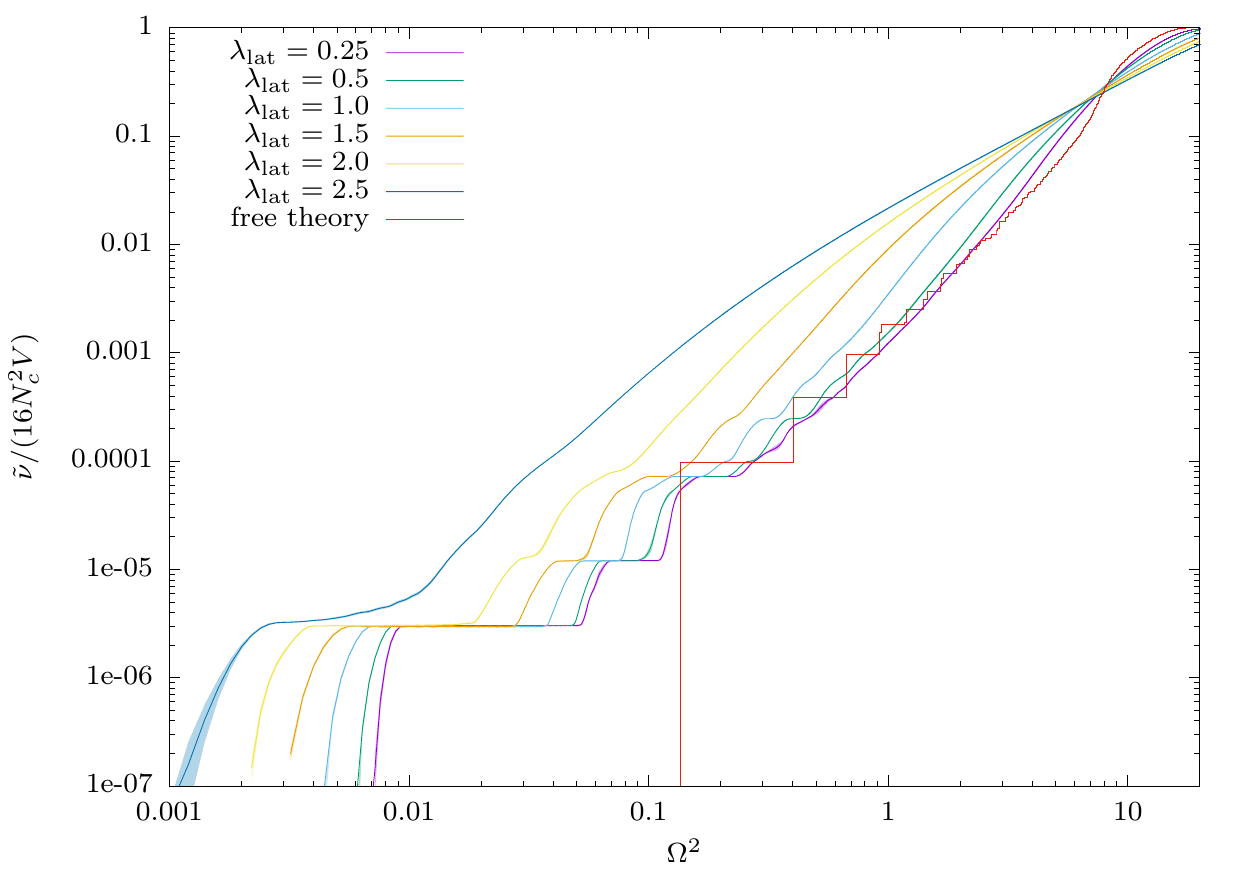}
  \caption{\label{fig:eig_mode}From \refcite{Bergner:2021ffz}, the normalized eigenmode number for $16^4$ lattices with gauge group U(2), including both the free theory (red) and the interacting theory with $0.25 \leq \lalat \leq 2.5$, plotted vs.\ the energy scale $\Om^2$ on log--log axes.}
\end{figure}

Figure~\ref{fig:eig_mode}, from \refcite{Bergner:2021ffz}, presents results for the normalized eigenmode number $\nu(\Om^2)$ from $16^4$ lattices with gauge group U(2) and a range of 't~Hooft couplings.
We include $\nu(\Om^2)$ for the free lattice theory, shown by the red line in this figure, which remains sensitive to the finite, discrete lattice space-time.
The smoother curves come from ensembles of field configurations generated with $0.25 \leq \lalat \leq 2.5$.

\begin{figure}[tbp]
  \centering
  \includegraphics[width=0.9\linewidth]{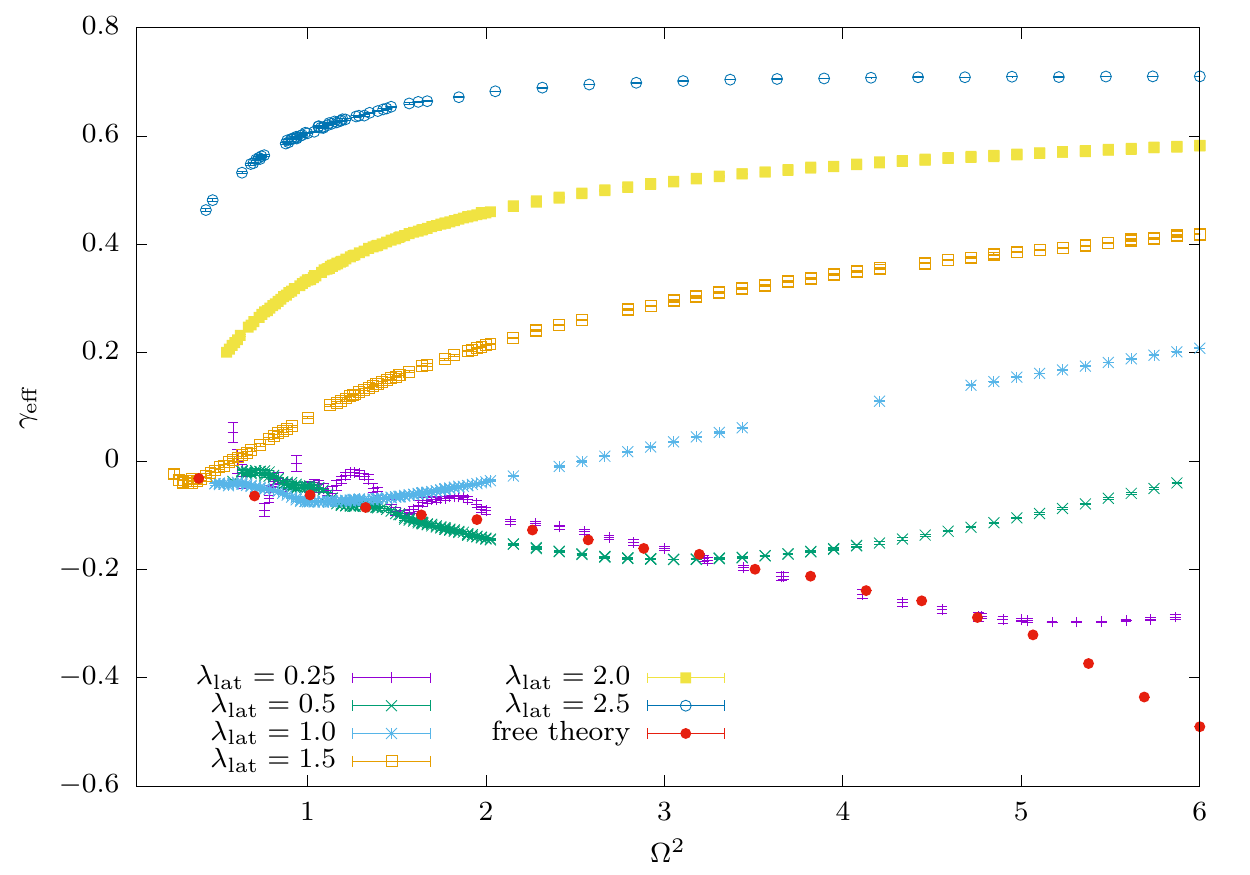}
  \caption{\label{fig:eig_gamma}From \refcite{Bergner:2021ffz}, the effective anomalous dimension \gaEff obtained from the U(2) $16^4$ results shown in \fig{fig:eig_mode}, plotted vs.\ the energy scale $\Om^2$ on linear axes.}
\end{figure}

According to \eq{eq:mode}, the slope of $\nu(\Om^2)$ on the log--log axes of \fig{fig:eig_mode} is directly related to the mass anomalous dimension $\ga_*$.
Due to the explicit breaking of conformality, we are only able to access a scale-dependent effective anomalous dimension $\gaEff(\Om^2)$, which we extract by fitting our data to \eq{eq:mode} in windows $\left[\Om^2, \Om^2 + \ell\right]$ with $\ell \in [0.03, 1]$ fixed for each ensemble.
The U(2) $16^4$ results for $\gaEff(\Om^2)$ from these fits are collected in \fig{fig:eig_gamma}.
Similar results for larger gauge groups and different lattice volumes are presented in \refcite{Bergner:2021ffz}.
From this figure we can see that the true $\ga_* = 0$ is recovered in the IR, for $\Om^2 \ll 1$.
We can also see that even the free theory on a $16^4$ lattice suffers from significant lattice artifacts as $\Om^2$ increases.
Finally, these artifacts clearly increase rapidly as the lattice 't~Hooft coupling becomes stronger, quantifying the challenge of successfully recovering the superconformal continuum field theory.

\section{\label{sec:Konishi}Konishi and SUGRA scaling dimensions} 
While the expectation $\ga_* = 0$ for the mass anomalous dimension discussed above makes it a useful tool with which to explore the breaking of supersymmetry and conformality in lattice studies of $\cN = 4$ SYM, it's also important to pursue non-trivial scaling dimensions using lattice field theory.
A compelling target is the scaling dimension of the `Konishi' operator
\begin{equation}
  \cO_K(x) = \sum_I \Tr\left[\Phi^I(x) \Phi^I(x)\right],
\end{equation}
where $\Phi^I$ are the six real scalar fields of the continuum theory.
This is the simplest conformal primary operator of $\cN = 4$ SYM, with a non-trivial scaling dimension $\De_K(\la) = 2 + \ga_K(\la)$ that has been predicted with weak-coupling perturbation theory~\cite{Fiamberti:2008sh, Bajnok:2008bm, Velizhanin:2008jd}, from holography at strong couplings $\la \to \infty$ with $\la \ll N$~\cite{Gubser:1998bc}, and for all couplings in the $N = \infty$ planar limit~\cite{Gromov:2009zb}.
Thanks to S-duality, which predicts an invariant spectrum of $\cN = 4$ SYM anomalous dimensions under the interchange $\frac{4\pi N}{\la} \llra \frac{\la}{4\pi N}$, the perturbative results are also relevant in the alternate strong-coupling regime $\la \gg N$~\cite{Beem:2013hha}.
In addition, the superconformal bootstrap program has been applied to analyze the Konishi anomalous dimension, with initial bounds on the maximum value $\ga_K$ could reach across all $\la$~\cite{Beem:2013qxa, Beem:2016wfs} recently being generalized to $\la$-dependent constraints~\cite{Chester:2021aun}. 

\begin{figure}[tbp]
  \centering
  \includegraphics[width=0.7\linewidth]{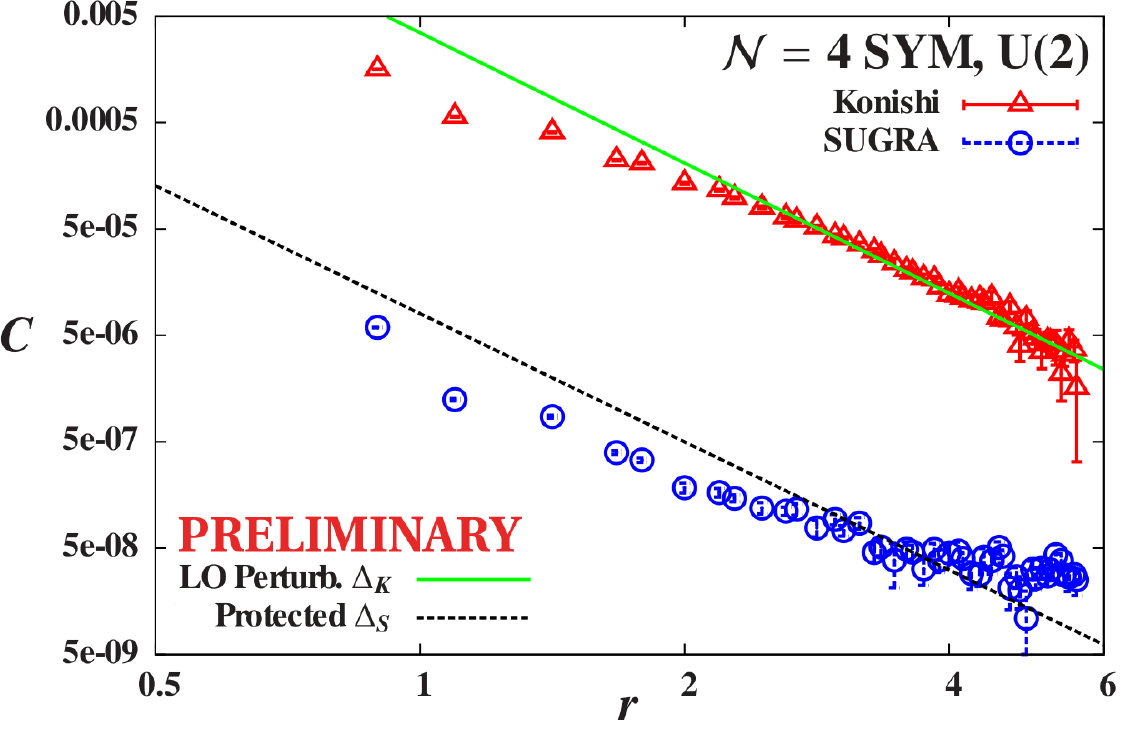}
  \caption{\label{fig:correlators}The Konishi and SUGRA two-point correlators for a U(2) $16^4$ ensemble with $\lalat = 0.5$, plotted vs.\ the distance $r$ on log--log axes.  The solid and dotted lines are not fits, but expectations based on leading-order perturbation theory and the exact value $\De_S = 2$, respectively.}
\end{figure}

In order to analyze the Konishi scaling dimension using our ensembles of lattice field configurations, we first need to isolate scalar fields $\varphi_a(n)$ that have been twisted into the complexified gauge links.
We do so using a polar decomposition,
\begin{equation}
  \cU_a(n) = e^{\varphi_a(n)} U_a(n).
\end{equation}
Because the Konishi operator is a scalar under the twisted rotation group, it picks up a non-zero vacuum expectation value (vev) that needs to be subtracted,
\begin{equation}
  \label{eq:Konishi}
  \cO_K^{\text{lat}}(n) = \sum_a \Tr\left[\varphi_a(n) \varphi_a(n)\right] - \mbox{vev}.
\end{equation}
Figure~\ref{fig:correlators} shows preliminary results for the Konishi correlator
\begin{equation}
  C_K(r) = \cO_K^{\text{lat}}(n + r)\cO_K^{\text{lat}}(n) \propto r^{-2\De_K},
\end{equation}
along with the corresponding correlator of the `SUGRA' operator
\begin{align}
                            \cO_S^{IJ}(x) & = \Tr\left[X^{\{I}(x) X^{J\}}(x)\right] &
  \left[\cO_S^{\text{lat}}\right]^{ab}(n) & = \Tr\left[X^{\{a}(n) X^{b\}}(n)\right].
\end{align}
In the continuum theory, $\cO_S$ transforms in the symmetric traceless $20'$ representation of the SO(6) R-symmetry, and it has no anomalous dimension: $\De_S = 2$ for all 't~Hooft couplings.
Considering gauge group U(2) on a $16^4$ lattice volume with $\lalat = 0.5$, the figure shows a range of $r$ in which the Konishi correlator follows a clear power law --- a straight line on these log--log axes, whose slope agrees with the leading-order perturbative prediction for $\De_K(\la)$.
The SUGRA correlator is noisier, due to being $\cO(100)$ times smaller in magnitude, but behaves consistently with its protected scaling dimension $\De_S = 2$.

Work remains in progress to quantify the systematic uncertainties that come into play when extracting $\De_K$ from correlators of this sort, which include sensitivity to the power-law fit range and the subtraction of the vev in \eq{eq:Konishi}, along with finite-volume and discretization artifacts.
So far we have had more success determining $\De_K$ through Monte Carlo renormalization group (MCRG) stability matrix analyses~\cite{Swendsen:1979gn}.
Treating the lattice system as a formally infinite sum of operators, $\sum_i c_i \cO_i$, essentially all of which are irrelevant in the RG sense, an RG blocking step $R_b$ defines a new system
\begin{equation}
  H^{(n)} = R_b H^{(n - 1)} = \sum_i c_i^{(n)} \cO_i^{(n)}
\end{equation}
based on $n$-times blocked operators $\cO_i^{(n)}$ with RG-flowed coefficients $c_i^{(n)}$.

For a conformal system like $\cN = 4$ SYM at any value of the 't~Hooft coupling $\la$, the conformal fixed point is characterized by $H^* = R_b H^*$ with couplings $c_i^*$.
Linearizing around this fixed point,
\begin{equation}
  c_i^{(n)} - c_i^* = \sum_k \left. \pderiv{c_i^{(n)}}{c_k^{(n - 1)}}\right|_{H^*}\left(c_k^{(n - 1)} - c_k^*\right) \equiv \sum_k T_{ik}^* \left(c_k^{(n - 1)} - c_k^*\right),
\end{equation}
defines the stability matrix $T_{ik}^*$.
In lattice calculations, the matrix elements of $T_{ik}^*$ come from correlators of the operators $\cO_i$ and $\cO_k$ evaluated after different numbers of RG blocking steps, while its eigenvalues provide the corresponding scaling dimensions.

\begin{figure}[tbp]
  \centering
  \includegraphics[width=0.7\linewidth]{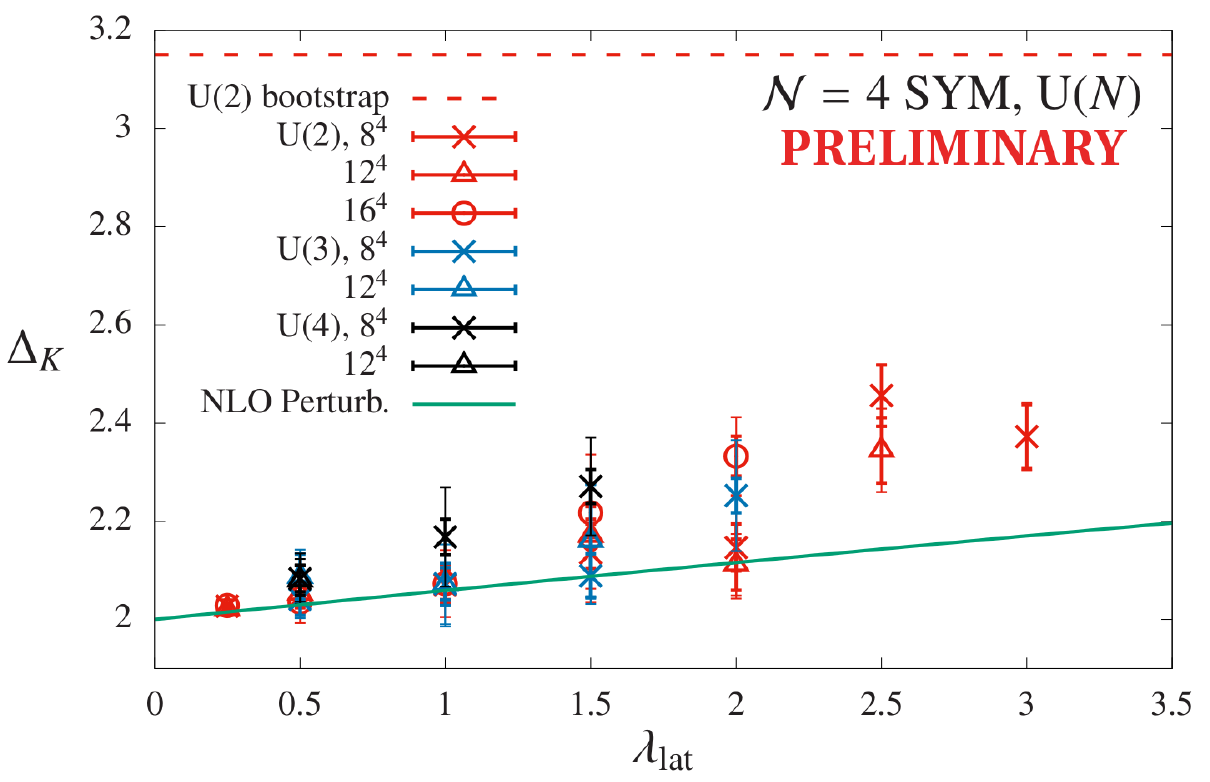}
  \caption{\label{fig:MCRG}Preliminary results for the Konishi scaling dimension $\De_K(\lalat)$ from MCRG analyses for gauge groups up to U(4) (different colors) and lattice volumes up to $16^4$ (different symbols).  The thin and thick error bars on each point correspond to statistical and total uncertainties, respectively.  The dashed line at the top of the plot is an upper bound from conformal bootstrap analyses~\cite{Beem:2013qxa}.}
\end{figure}

The lattice $\cN = 4$ SYM RG blocking transformation that we use was introduced by \refcite{Catterall:2014mha}.
Figure~\ref{fig:MCRG} presents our preliminary MCRG results for the Konishi scaling dimension $\De_K(\lalat)$, considering gauge groups U(2), U(3) and U(4) along with lattice volumes up to $16^4$.
In these analyses, we use both $\cO_K^{\text{lat}}$ and $\cO_S^{\text{lat}}$ to construct the stability matrix, applying APE-like smearing to expand the basis of operators.
We have also imposed $\De_S = 2$ as a constraint, which ensures physical results for $\De_K \geq 2$ corresponding to $\ga_K \geq 0$.
By varying the operator basis in the analyses we have obtained initial estimates for systematic uncertainties.
The thin error bars on each point combine statistical and systematic uncertainties in quadrature, while the think error bars show the statistical uncertainties themselves.

For the range of 't~Hooft couplings we have analyzed so far, $\lalat \leq 3$, our results for the Konishi scaling dimension are consistent with perturbation theory, and well below the U(2) bootstrap bound on the maximum value of $\De_K$ across all $\la$~\cite{Beem:2013qxa}.
This is not surprising, given that the perturbative expansion parameter is $\la / 4\pi^2$.
It is a longstanding challenge to reach stronger couplings~\cite{Catterall:2020lsi, Schaich:2022xgy}, and another challenge comes from the twisted formulation we employ.
Because the twisted rotation group involves only an $\SO{4}_R$ subgroup of the full $\SO{6}_R$ R-symmetry, the SUGRA operator $\cO_S$ is projected from the $20'$ representation of SO(6) to a collection of SO(4) representations.
In particular, these include an $\SO{4}_R$-singlet piece that mixes with our lattice Konishi operator $\cO_K^{\text{lat}}$ and may require variational analyses to disentangle.

\section{\label{sec:conc}Conclusions and next steps} 
This proceedings has explored the role of broken conformality in numerical lattice field theory analyses of $\cN = 4$ SYM by considering two conformal scaling dimensions.
The first, a mass anomalous dimension related to the eigenmode number of the fermion operator, is expected to vanish, $\ga_* = 0$, for all values of $\la$, making it a useful tool with which to quantify the effects of broken conformality.
By stochastically reconstructing the Chebyshev expansion of the spectral density, we can extract a scale-dependent effective anomalous dimension that correctly converges to the true $\ga_* = 0$ in the IR.
Even the free theory exhibits significant lattice artifacts at higher energy scales, which increase rapidly as the lattice 't~Hooft coupling \lalat becomes stronger.

Second, turning to the non-trivial, $\la$-dependent scaling dimension of the Konishi operator, we have been able to verify conformal power-law scaling of the Konishi correlator with an appropriate value of $\De_K$.
Further analyses using the MCRG stability matrix method have produced preliminary results for the scaling dimension that are consistent with continuum perturbation theory for $\lalat \leq 3$.
Work is in progress to finalize these results, and in the longer term to push to stronger 't~Hooft couplings, with particular interest in the self-dual point $\la = 4\pi N$.
We are also investigating potential applications of the gradient flow to lattice $\cN = 4$ SYM.
These could include non-perturbatively testing the extent to which broken conformality results in a non-vanishing $\beta$-function, as well as novel analyses of conformal scaling dimensions as proposed by \refcite{Carosso:2018bmz}.

\vspace{20 pt} 
\noindent \textsc{Acknowledgments:}~I thank Georg Bergner, Simon Catterall and Joel Giedt for collaboration on the investigations summarized here.
This work was supported by UK Research and Innovation Future Leader Fellowship {MR/S015418/1} and STFC grant {ST/T000988/1}.
Numerical calculations were carried out at the University of Liverpool, the University of Bern, and on USQCD facilities at Fermilab funded by the US Department of Energy.

\bibliographystyle{JHEP}
\bibliography{lattice22}
\end{document}